# On modelling of bioavailability of drugs in terms of conservation laws

S.Piekarski (1), M.Rewekant (2)
Institute of Fundamental Technological Research Polish Academy of Sciences (1), Medical University of Warsaw, Poland

**Abstract**

In the text S.Piekarski, M.Rewekant, "On applications of conservation laws in pharmacokinetics",(arXiv:1208.3847) it has been mentioned that some information on bioavailability and bioequivalence of drugs can be obtained from simulations based on the conservation laws. Here we shortly discuss that possibility starting from the fundamental pharmacokinetic parameter called AUC (Area Under the Curve). The curve is is the profile shape of plasma drug concentration in time intervals after drug administration into organism. Our aim here is to give some information on the subject for the reader with no experience in pharmacokinetics.

# Introduction

Bioavailability of a drug product is usually defined in terms of rate and extent to which the active drug ingredient is absorbed and becomes available at the site of drug action. Two drug formulations are said to be bioequivalent if their bioavailabilities are statistically similar. In pharmacokinetics, one usually measures the profiles of the total drug concentration in the circulating blood (that is, in a "central compartment"). "Profile" means here concentration – time curve. The total drug concentration is a sum of the concentration of a drug bound to blood proteins and the concentration of a free form of a drug. Several approaches were proposed in the literature to estimate the bioavailability of a drug from the profile shape. Some of them define functions depending on a single profile and then different profiles are compared indirectly, after comparing functions of a single profile. An alternative approach is to define an appropriate distance measure between two profiles. Such distance measures are called bioequivalence metrics and the corresponding distance estimates the similarity or dissimilarity of the two profiles. In general, functions of a single profile are called pharmacokinetic (PK) parameters and the main PK parameters are: AUC (area under curve), $C_{max}$ (maximum drug plasma concentration), $t_{max}$ (time to the maximum concentration after drug administration), $AUC_p$ (partial area under the profile shape from zero to $t_{max}$). It is worth to mention an option suggested by United States Food and Drug Administration (FDA). It is formulated in terms of functions of a single profile and suggest the following criteria for the assessment of bioequivalence between test (T) and reference (R) drug formulations: test and reference formulations are bioequivalent if boundaries of 90% confidence intervals for each of $AUC_T/AUC_R$ and $\frac{C_{max}^T}{C_{max}^R}$ lie between 0.8 and 1.25. However, it is worth to stress that the concentration profile is a result of different processes taking place in patient's organism, shortly named ADME; Absorption, Distribution, Metabolism and Excretion of a drug. In order to take into account all these processes explicitly one should write the corresponding evolutionary system of conservation equations (derived from the balance laws). Then it should be possible to compute all "pharmacokinetic parameters" and "distance metrics" from the initial and boundary conditions (under the assumption, that the explicit form of evolution equation is known). Obviously, in order to obtain even approximate form of such equations one would need much more experiments than measuring a total concentration of drug in central compartment. There is an additional reason for considering evolution equations here: there are different version of pharmacokinetics and there are inconsistencies between them. After using mathematical simulations derived from conservation laws, the reasons for those inconsistencies should become more transparent. The above remarks are very general and in order to give a simple example one can take the following problem. In standard pharmacokinetics, one usually identifies the quantity of a drug administered into the patient's organism with AUC. However, comparison with the simplest models of ADME processes shows that such identification is not obvious.

In order to check how good (or how bad) is a such approximation one can take the simplest possible model of a single intravenous administration of a drug since in this case the full information on the quantity of a drug introduced into the patient's organism is contained in the initial condition. For a given initial condition, one can compute AUC and compare AUC with the information contained in initial condition.

## Conclusions

A rigorous discussion should include the discussion of parameters of evolution equations and its relations with quantities known from standard pharmacokinetics. For such a discussion, one needs a lot of information on the history of pharmacokinetics and more mathematical models (for example, that concerning stationary state and volume of distribution).

As it has been already mentioned, Aldo Rescigno raised a lot of criticism against the standard formulation of pharmacokinetics and also here the natural reaction to that criticism is to comment it from the point of view of the theory of conservation laws. Finally, it is not excluded that the definition of "bioavailability" can be defined more precisely (in terms of differencies between equations modeling both compared drugs). We hope to write about all that later.